\documentclass[lettersize,journal]{IEEEtran}
\usepackage{amsmath,amsfonts,amssymb}
\usepackage{amsthm}
\usepackage{algorithmic}
\usepackage{algorithm}
\usepackage{array}
\usepackage[caption=false,font=footnotesize]{subfig}
\usepackage{textcomp}
\usepackage{stfloats}
\usepackage{url}
\usepackage{verbatim}
\usepackage{graphicx}
\usepackage{cite}
\usepackage{bm}
\usepackage{color}
\usepackage{multirow}
\newcommand{\bbR}{\mathbb{R}}

\theoremstyle{definition}
\newtheorem{theorem}{Theorem}

\newtheorem{lemma}{Lemma}
\newtheorem{rem}{Remark}
\newtheorem{col}{Corollary}

\newtheorem{ass}{Assumption}

\newcommand{\bbZ}{\mathbb{Z}}
\newcommand{\bbE}{\mathbb{E}}
\newcommand{\bbV}{\mathbb{V}}
\newcommand{\bbP}{\mathbb{P}}

\newcommand{\hatxi}{\widehat{\xi}}
\newcommand{\hatv}{\widehat{v}}
\newcommand{\hatA}{\widehat{A}}
\newcommand{\hatB}{\widehat{B}}
\newcommand{\hatc}{\widehat{c}}
\newcommand{\hattheta}{\widehat{\theta}}
\newcommand{\hatgamma}{\widehat{\gamma}}

\newcommand{\barxi}{\overline{\xi}}
\newcommand{\ubarxi}{\underline{\xi}}
\newcommand{\barv}{\overline{v}}
\newcommand{\ubarv}{\underline{v}}
\newcommand{\bardelta}{\overline{\delta}}
\newcommand{\barx}{\overline{x}}
\newcommand{\ubarx}{\underline{x}}
\newcommand{\baru}{\overline{u}}
\newcommand{\ubaru}{\underline{u}}
\newcommand{\bartau}{\overline{\tau}}
\newcommand{\ubartau}{\underline{\tau}}
\newcommand{\barQ}{\overline{Q}}
\newcommand{\ubarQ}{\underline{Q}}
\newcommand{\barP}{\overline{P}}
\newcommand{\ubarP}{\underline{P}}

\newcommand{\bfone}{\mathbf{1}}

\newcommand{\st}{\mathrm{subject\ to\ }}

\newcommand{\conv}{\mathrm{conv}}
\newcommand{\soft}{\mathrm{soft}}
\newcommand{\req}{\mathrm{req}}

\newcommand{\eng}{\mathrm{eng}}
\newcommand{\mot}{\mathrm{mot}}
\newcommand{\brk}{\mathrm{brk}}
\newcommand{\cmd}{\mathrm{cmd}}
\newcommand{\rmds}{\mathrm{ds}}
\newcommand{\chg}{\mathrm{chg}}
\newcommand{\target}{\mathrm{target}}

\begin{document}

\title{Convex Chance-Constrained Stochastic Control under Uncertain Specifications with Application to Learning-Based Hybrid Powertrain Control}
\author{Teruki Kato, Ryotaro Shima, and Kenji Kashima,~\IEEEmembership{Senior Member,~IEEE}%
% \thanks{Manuscript submitted January 23, 2026}
\thanks{This work was not supported by any external funding.}
\thanks{T. Kato and R. Shima are with Toyota Central R\&D Labs., Inc., Nagakute, Aichi, Japan
(e-mail: \mbox{teruki.kato.tg@mosk.tytlabs.co.jp}; \mbox{ryotaro.shima@mosk.tytlabs.co.jp}).}%
\thanks{K. Kashima is with the Graduate School of Informatics, Kyoto University, Kyoto, Japan
(e-mail: \mbox{kk@i.kyoto-u.ac.jp}).}%
}
\maketitle

\begin{abstract}
This paper proposes a strictly convex chance-constrained stochastic control framework that accounts for uncertainty in control specifications such as reference trajectories and operational constraints.
Control inputs and risk allocation are jointly optimized under general, possibly non-Gaussian, uncertainties through sufficient deterministic conditions for probabilistic constraint satisfaction.
A decreasing risk regularizer ensures existence and full use of the allowable risk.
Under the stated regularity assumptions, the joint solution is unique and locally Lipschitz-continuous, supporting stable numerical optimization.
The formulation is extended to learning-based exactly linearizable (EL) models through monotone state and input transformations.
A hybrid powertrain example compares deterministic control, equal risk allocation, and optimized risk allocation with and without regularization in terms of charge sustaining, constraint violations, closed-loop cost, and solver time.
\end{abstract}


\section{Introduction}
In real-world control systems, stochastic uncertainties arise in modeling errors, physical parameters, and measurements.
Although stochastic control methods address reliability under such uncertainties \cite{smpc_review_farina}, uncertainty can also affect control specifications, including objective functions and constraints, through external environments, human decision-making, and statistical models.
For example, automotive powertrain control \cite{vehicle_propulsion_systems} must track the requested vehicle speed while satisfying constraints on exhaust emissions and battery charge.
The requested speed is uncertain because of driver decision-making \cite{smpc_hev,longitudinal_sde}, probabilistic planning in automated driving \cite{diffuser,diffusion_planner}, and surrounding traffic and road conditions.
Consequently, both reference trajectories and operating constraints can be uncertain.

Chance-constrained control limits the probability of constraint violation, referred to as risk.
Since joint chance constraints are difficult to handle directly, a common approach is to allocate a total allowable risk among individual constraints and replace the resulting constraints with tractable deterministic conditions.
For linear systems with general additive uncertainties and fixed disturbance-feedback gains, joint optimization of control inputs and risk allocation can be convex \cite{braatz_ijc2020}.
Gaussian-process models enable stochastic control of learned nonlinear dynamics \cite{cautious_gpr_mpc}, while iterative risk allocation offers an alternative approach to joint chance constraints \cite{ira}.
These methods do not establish strict convexity of the joint optimization over control inputs and risk allocation.
Strict convexity is important for uniqueness and, under suitable regularity assumptions, continuity of the optimizer \cite{boyd_convex,regularity_opt}; it also supports numerical stability.

Methods addressing uncertain environments and obstacles use chance constraints with Gaussian uncertainty \cite{uncertain_env_cairano_nonlin,uncertain_obs_multi,uncertain_obs_cbf}.
Scenario-based hybrid powertrain control accounts for uncertain speed-dependent input bounds and provides a strictly convex formulation but does not impose chance constraints \cite{east_cannon_smpc_hev}.
Table~\ref{tab:intro} summarizes these distinctions.
For learned nonlinear models \cite{brunton_kutz}, deterministic convex model predictive control (MPC) can be developed using either exactly linearizable (EL) models \cite{exlin,deep_sign_definite} or input convex neural networks (ICNNs) \cite{icnn_control}.
This study uses the EL approach and extends it to uncertain specifications.

\begin{table*}[!t]
    \caption{Comparison with existing studies}\label{tab:intro}
    \centering
    \small
    \begin{tabular*}{\textwidth}{@{\extracolsep{\fill}}lccccc@{}}
        \hline
        Method & \shortstack{Specification\\uncertainty} & \shortstack{Chance\\constraints} & Non-Gaussian & \shortstack{Strict\\convexity} & \shortstack{Nonlinear\\dynamics} \\
        \hline
        \cite{braatz_ijc2020} & $\times$ & \checkmark & \checkmark & $\times$ & $\times$ \\
        \cite{cautious_gpr_mpc} & $\times$ & \checkmark & $\times$ & $\times$ & \checkmark \\
        \cite{ira} & $\times$ & \checkmark & $\times$ & $\times$ & $\times$ \\
        \cite{smpc_hev} & $\times$ & $\times$ & \checkmark & \checkmark & $\times$ \\
        \cite{uncertain_env_cairano_nonlin,uncertain_obs_multi,uncertain_obs_cbf} & \checkmark & \checkmark & $\times$ & $\times$ & \checkmark \\
        \cite{east_cannon_smpc_hev} & \checkmark & $\times$ & \checkmark & \checkmark & $\times$ \\
        This paper (proposed) & \checkmark & \checkmark & \checkmark & \checkmark & \checkmark\ (EL models) \\
        \hline
    \end{tabular*}
\end{table*}

The contributions of the study are:
\begin{itemize}
    \item A strictly convex formulation that jointly optimizes control inputs and risk allocation under uncertain references and constraints.
    \item A risk regularizer that ensures existence, uniqueness, and continuity, as well as full use of the allowable risk, under the stated assumptions.
    \item An extension to learned EL models, validated through hybrid powertrain MPC.
    % \cite{mpc_handbook}
\end{itemize}

The notation used in this paper is defined as follows.
The set of nonnegative integers is denoted by $\bbZ_\geq$.
For a positive integer $N$, $\bbZ_N$ denotes the set $\{1,\ldots,N\}$.
Elementwise inequalities between vectors are denoted by $\leq$, and the elementwise sum of squares is denoted by $\| \cdot \|^2$.
The all-ones vector is denoted by $\mathbf{1}$.
For a random variable, expectation and variance are denoted by $\bbE[\cdot]$ and $\bbV[\cdot]$, respectively, and the probability of an event is denoted by $\bbP[\cdot]$.

\section{Preliminaries: Deterministic control for a linear system}
In this section, we consider the following linear system:
\begin{align}
    \xi_{k+1} = A\xi_k + Bv_k + c(\theta_k) \label{eq:state_eq_lin}
\end{align}
Here $k\in\bbZ_\geq$, $\xi_k \in \bbR^{n_\xi}$, $v_k \in \bbR^{n_v}$, $\theta_k \in \bbR^{n_\theta}$, and $c:\bbR^{n_\theta} \to \bbR^{n_\xi}$.
The parameter $\theta_k$ is a generic exogenous variable representing disturbances, model mismatch, and uncertain quantities that define the reference and constraints.
In this section, $\theta$ is assumed to be deterministic and known.
The optimal control problem for system \eqref{eq:state_eq_lin} is formulated as follows:
\begin{align}
    &\min_{v,\gamma}\ \sum_{k=0}^{N-1}\Bigl[
    w^\req\|\xi_{k+1}-\xi^\req(\theta_{k+1})\|^2
    +w^v\|v_k\|^2 \nonumber\\
    &\hspace{43mm}+w^\soft\|\gamma_{k+1}\|^2\Bigr] \label{eq:opt_control_lin_obj}\\
    &\st \nonumber\\
    &\xi_{k+1}=A\xi_k+Bv_k+c(\theta_k)\\
    &\ubarxi(\theta_{k+1})\leq\xi_{k+1}\leq\barxi(\theta_{k+1})\\
    &\ubarv(\theta_k)\leq v_k\leq\barv(\theta_k)\\
    &\xi_{k+1}-\barxi^\soft(\theta_{k+1})\leq\gamma_{k+1},\quad 0\leq\gamma_{k+1}
    \label{eq:opt_control_lin_const}\\
    &k=0,\ldots,N-1 \nonumber
\end{align}
The nonnegative slack $\gamma_{k+1}$ makes the last term equivalent to the penalty $w^\soft\|\max(\xi_{k+1}-\barxi^\soft(\theta_{k+1}),0)\|^2$ \cite{boyd_convex}.
The weights $w^\req,w^v,w^\soft$ are positive and penalize tracking error, control effort, and soft-constraint violations, respectively.
Lower soft bounds are addressed by changing signs.
The dependence of the reference and bounds on $\theta$ allows both dynamics and control specifications to depend on this parameter.
The above problem is a strongly convex quadratic program after eliminating the state variables.

Let the stacked vectors be $\hatxi:=[\xi_1^T,\ldots,\xi_N^T]^T$, $\hatv:=[v_0^T,\ldots,v_{N-1}^T]^T$, and $\hattheta:=[\theta_0^T,\ldots,\theta_N^T]^T$.
Then \eqref{eq:state_eq_lin} gives
\begin{align}
    \hatxi=\hatA\xi_0+\hatB\hatv+\hatc(\hattheta).\label{eq:stacked_state}
\end{align}
The block elements, for $i,j\in\bbZ_N$, are
\begin{align}
    &\hatA_i=A^i,\qquad
    \hatB_{ij}=\begin{cases}A^{i-j}B,&j\leq i,\\0,&j>i,\end{cases}\nonumber\\
    &\hatc_i(\hattheta)=\sum_{\ell=0}^{i-1}A^{i-1-\ell}c(\theta_\ell).
    \label{eq:stacked_blocks}
\end{align}
Define $\hatgamma:=[\gamma_1^T,\ldots,\gamma_N^T]^T$ and stack the reference and state bounds over $1,\ldots,N$, and the input bounds over $0,\ldots,N-1$.
With $n_X:=3Nn_\xi+2Nn_v$, set
\begin{align}
    X(\hattheta)&:=
    \begin{bmatrix}
        \ubarxi(\hattheta)-\hatc(\hattheta)\\
        \hatc(\hattheta)-\barxi(\hattheta)\\
        \ubarv(\hattheta)\\
        -\barv(\hattheta)\\
        \hatc(\hattheta)-\barxi^\soft(\hattheta)
    \end{bmatrix}\in\bbR^{n_X},\label{eq:X_def}\\
    y(\hatv,\hatgamma)&:=
    \begin{bmatrix}
        \hatA\xi_0+\hatB\hatv\\
        -\hatA\xi_0-\hatB\hatv\\
        \hatv\\-\hatv\\
        \hatgamma-\hatA\xi_0-\hatB\hatv
    \end{bmatrix}\in\bbR^{n_X}.\label{eq:y_def}
\end{align}
Ignoring objective terms independent of the decision variables, the compact problem is
\begin{align}
    &\min_{\hatv,\hatgamma\geq0}\
    w^\req\|\hatB\hatv\|^2+w^v\|\hatv\|^2+w^\soft\|\hatgamma\|^2\nonumber\\
    &\hspace{8mm}+2w^\req[\hatA\xi_0+\hatc(\hattheta)-\xi^\req(\hattheta)]^T\hatB\hatv
    \label{eq:opt_control_hat_obj}\\
    &\st X(\hattheta)\leq y(\hatv,\hatgamma).
    \label{eq:opt_control_hat_const}
\end{align}

\section{Stochastic control accounting for uncertainty in control specifications for a linear system}
When $\theta$ is random, Problem~\eqref{eq:opt_control_hat_obj}--\eqref{eq:opt_control_hat_const} becomes a stochastic control problem with uncertain specifications.
We use the following assumption to derive a convex sufficient condition for its chance constraints.
\begin{ass}\label{ass:theta_distribution}
One of the following holds:
\begin{enumerate}
    \item[(mv)] The mean and variance of $X_j(\hattheta)$ $(j \in \bbZ_{n_X})$ are known.
    \item[(bd)] The mean of $X_j(\hattheta)$ $(j \in \bbZ_{n_X})$ is known, and there exist $L^X, U^X \in \bbR^{n_X}$ such that $X_j(\hattheta)\in[L_j^X, U_j^X]$ $(j \in \bbZ_{n_X})$ almost surely.
    \item[(cdf)] The cumulative distribution function $F_{X_j}$ of $X_j(\hattheta)$ $(j \in \bbZ_{n_X})$ and its inverse $F_{X_j}^{-1}$ are known.
    Furthermore, there exists $x_j^*\in\bbR$ $(j \in \bbZ_{n_X})$ such that the probability density function $f_{X_j}$ of $X_j(\hattheta)$ is nonzero and monotonically decreasing on $[x_j^*, \infty)$.
\end{enumerate}
\end{ass}
Under Assumption~\ref{ass:theta_distribution}, define the expected objective, omitting constant terms, as follows:
\begin{align}
    J(\hatv,\hatgamma):={}&w^\req\|\hatB\hatv\|^2+w^v\|\hatv\|^2+w^\soft\|\hatgamma\|^2\nonumber\\
    &+2w^\req\bbE[\hatA\xi_0+\hatc(\hattheta)-\xi^\req(\hattheta)]^T\hatB\hatv.
    \label{eq:expected_cost}
\end{align}
The chance-constrained counterpart of Problem~\eqref{eq:opt_control_hat_obj}--\eqref{eq:opt_control_hat_const} is
\begin{align}
    &\inf_{\hatv,\hatgamma\geq0}\ J(\hatv,\hatgamma)\label{eq:joint_chance_obj}\\
    &\st\bbP[X(\hattheta)\leq y(\hatv,\hatgamma)]\geq1-\bardelta.
    \label{eq:joint_chance_const}
\end{align}
Here $\bardelta\in(0,1)$ denotes the total allowable risk.
We use $\inf$ rather than $\min$ because the existence of an optimizer is not guaranteed at this point.
A sufficient condition for the joint chance constraint follows from Boole's inequality (Corollary~\ref{cor:boole} in the Appendix), yielding
\begin{align}
    &\inf_{\hatv,\hatgamma\geq0,\,0<\delta<1}\ J(\hatv,\hatgamma)+w^\delta r(\delta)
    \label{eq:risk_joint_bool_obj}\\
    &\st\bbP[X_j(\hattheta)\leq y_j(\hatv,\hatgamma)]\geq1-\delta_j
    \quad(j\in\bbZ_{n_X})\label{eq:risk_joint_bool_chance}\\
    &\bfone^T\delta\leq\bardelta.\label{eq:risk_joint_bool_bardelta}
\end{align}
Here $\delta\in\bbR^{n_X}$ is the risk-allocation variable, $r:(0,1)^{n_X}\to\bbR$ is a $C^2$ regularizer, and $w^\delta>0$ is its weight.
To obtain tractable deterministic constraints, consider
\begin{align}
    &\inf_{\hatv,\hatgamma\geq0,\,0<\delta<1}\ J(\hatv,\hatgamma)+w^\delta r(\delta)
    \label{eq:risk_joint_convex_obj}\\
    &\st\psi_j(\delta_j)\leq y_j(\hatv,\hatgamma)\quad(j\in\bbZ_{n_X})
    \label{eq:risk_joint_convex_chance}\\
    &\bfone^T\delta\leq\bardelta.\label{eq:risk_joint_convex_bardelta}
\end{align}
Here the functions $\psi_j:(0,1)\to\bbR$ $(j \in \bbZ_{n_X})$ are defined, depending on Assumption~\ref{ass:theta_distribution}, as follows:
\begin{align}
    &\psi_j(\delta_j) :=
    \begin{cases}
        \bbE\left[ X_j(\hattheta) \right] + \sqrt{ \dfrac{1-\delta_j}{\delta_j} \bbV\left[ X_j(\hattheta) \right] } & \text{if (mv)}, \\
        \bbE\left[ X_j(\hattheta) \right] + \dfrac{U_j^X - L_j^X}{\sqrt{2}}\sqrt{ -\log(\delta_j) } & \text{if (bd)}, \\
        F_{X_j}^{-1}(1-\delta_j) & \text{if (cdf)}
    \end{cases}
    \label{eq:psi_def}
\end{align}
We also define the parameter $\delta^\conv\in(0,1)^{n_X}$ as follows:
\begin{align}
    &\delta_j^\conv :=
    \begin{cases}
        3/4 & \text{if (mv)}, \\
        e^{-1/2} & \text{if (bd)}, \\
        1-F_{X_j}(x_j^*) & \text{if (cdf)}
    \end{cases}
    \ (j \in \bbZ_{n_X})
    \label{eq:delta_conv_def}
\end{align}
In this case, regarding sufficiency and convexity of Problem~\eqref{eq:risk_joint_convex_obj}--\eqref{eq:risk_joint_convex_bardelta}, the following result holds.
\begin{theorem}\label{thm:risk_joint_convexity}
For the chance-constrained stochastic control problem
\eqref{eq:joint_chance_obj}--\eqref{eq:joint_chance_const},
which is a probabilistic counterpart of the deterministic optimal control problem
\eqref{eq:opt_control_lin_obj}--\eqref{eq:opt_control_lin_const},
the following statements hold under Assumption~\ref{ass:theta_distribution}.
\begin{enumerate}
    \item Any feasible solution of Problem~\eqref{eq:risk_joint_convex_obj}--\eqref{eq:risk_joint_convex_bardelta} is also a feasible solution of Problem \eqref{eq:joint_chance_obj}--\eqref{eq:joint_chance_const}.
    \item If the allowable risk satisfies $\bardelta\leq \delta_j^\conv$ $(j \in \bbZ_{n_X})$, then the feasible set of Problem~\eqref{eq:risk_joint_convex_obj}--\eqref{eq:risk_joint_convex_bardelta} is convex.
    Moreover, if the regularizer $r$ is strictly convex, then Problem \eqref{eq:risk_joint_convex_obj}--\eqref{eq:risk_joint_convex_bardelta} is strictly convex.
\end{enumerate}
\end{theorem}
\begin{proof}
First, we prove 1).
From the discussion above, any feasible solution of Problem~\eqref{eq:risk_joint_bool_obj}--\eqref{eq:risk_joint_bool_bardelta} is a feasible solution of Problem~\eqref{eq:joint_chance_obj}--\eqref{eq:joint_chance_const}.
Therefore, it suffices to show that any feasible solution of Problem~\eqref{eq:risk_joint_convex_obj}--\eqref{eq:risk_joint_convex_bardelta} is a feasible solution of Problem~\eqref{eq:risk_joint_bool_obj}--\eqref{eq:risk_joint_bool_bardelta}.
In case (mv), this follows from a corollary of Cantelli's inequality (Corollary~\ref{cor:cantelli}) given in the Appendix; in case (bd), it follows from a corollary of Hoeffding's inequality (Corollary~\ref{cor:hoeffding}) given in the Appendix; and in case (cdf), it follows from the definition of the cumulative distribution function.

Next, for item 2), the assumptions and \eqref{eq:risk_joint_convex_bardelta} imply $\delta\leq\delta^\conv$.
The second derivatives of the expressions in \eqref{eq:psi_def} show that each $\psi_j$ is convex on $(0,\delta_j^\conv]$; in case (cdf), this follows from the decreasing density on the specified tail.
Since $y$ is affine, the feasible set is convex.
The objective is strictly convex in $(\hatv,\hatgamma)$ because $w^v,w^\soft>0$; adding a strictly convex $r$ makes it strictly convex in all decision variables.
\end{proof}

\begin{rem}
In case (mv), each deterministic condition \eqref{eq:risk_joint_convex_chance} is equivalent to the individual distributionally robust chance constraint over distributions with the prescribed mean and variance \cite{cantelli_distributionally_robust}.
\end{rem}

To encourage full use of the allowable risk while preventing individual risk allocations from approaching zero, we introduce a decreasing regularizer that diverges at zero.

\begin{theorem}\label{thm:risk_joint_solution}
Assume that Problem \eqref{eq:risk_joint_convex_obj}--\eqref{eq:risk_joint_convex_bardelta} is feasible. 
Moreover, assume that the regularizer $r$ is separable as 
$r(\delta)=\sum_{j=1}^{n_X} r_j(\delta_j)$ with functions 
$r_j:(0,1)\to\bbR$ $(j\in\bbZ_{n_X})$ that are strictly monotonically decreasing and satisfy 
$\lim_{\delta_j\to+0} r_j(\delta_j)=+\infty$.
Then:
\begin{enumerate}
    \item Problem \eqref{eq:risk_joint_convex_obj}--\eqref{eq:risk_joint_convex_bardelta} admits an optimal solution.
    \item For an optimal solution $(\hatv^*, \hatgamma^*, \delta^*)$ of Problem \eqref{eq:risk_joint_convex_obj}--\eqref{eq:risk_joint_convex_bardelta}$,$ we have $\bfone^T \delta^* = \bardelta$.
    \item Assume $\bardelta\leq \delta_j^\conv$ $(j\in\bbZ_{n_X})$ and that $r$ is strictly convex.
    Moreover, assume that, for the optimal solution $(\hatv^*(\xi_0), \hatgamma^*(\xi_0), \delta^*(\xi_0))$ of Problem \eqref{eq:risk_joint_convex_obj}--\eqref{eq:risk_joint_convex_bardelta} parameterized by the initial state $\xi_0\in\bbR^{n_\xi}$, the linear independence constraint qualification holds.
    Then the optimal solution of Problem \eqref{eq:risk_joint_convex_obj}--\eqref{eq:risk_joint_convex_bardelta} is unique, and $(\hatv^*(\xi_0), \hatgamma^*(\xi_0), \delta^*(\xi_0))$ is locally Lipschitz-continuous with respect to $\xi_0$.
\end{enumerate}
\end{theorem}
\begin{proof}
We first prove 1).
By assumption, the objective diverges to $+\infty$ as $\delta_j\to +0$ $(j\in\bbZ_{n_X})$; therefore, for a sufficiently small positive number $\varepsilon$, 
we can, without loss of equivalence, restrict the feasible set by $\varepsilon\bfone \leq \delta \leq \bardelta\bfone$.
Moreover, since the objective diverges to $+\infty$ as $\|\hatv\|+\|\hatgamma\|\to\infty$, any nonempty sublevel set of the objective over the feasible set is bounded. Together with the above restriction on $\delta$, such a sublevel set is compact.
Because the objective is continuous on this compact set, Problem~\eqref{eq:risk_joint_convex_obj}--\eqref{eq:risk_joint_convex_bardelta} admits an optimal solution.

Next, we prove 2) by contradiction.
From item 1), an optimal solution $(\hatv^*, \hatgamma^*, \delta^*)$ exists; suppose that it satisfies $\bfone^T \delta^* < \bardelta$.
Here, since $\psi_j(\delta_j)$ in the constraint corresponding to the risk $\delta_j$ is monotonically decreasing 
(that is, increasing $\delta_j$ does not shrink the feasible set of the remaining variables), 
there exists a sufficiently small positive number $\varepsilon$ such that $(\hatv^*,\ \hatgamma^*,\ \delta^{**} := \delta^* + \varepsilon \bfone)$ is also feasible.
Together with the strict monotone decrease of $r_j$, this implies that $(\hatv^*,\ \hatgamma^*,\ \delta^{**})$ is a feasible solution with a strictly smaller objective value 
than $(\hatv^*,\ \hatgamma^*,\ \delta^*)$.
This contradicts the optimality of $(\hatv^*, \hatgamma^*, \delta^*)$.
Hence $\bfone^T \delta^* = \bardelta$ holds.

Finally, item 3) is proved.
Under the stated assumptions, the problem is strictly convex by Theorem~\ref{thm:risk_joint_convexity} and admits an optimal solution by item~1); hence, a unique optimal solution exists.
Moreover, since the objective and constraint functions are $C^2$ and the linear independence constraint qualification holds, the optimal solution is locally Lipschitz-continuous with respect to $\xi_0$ \cite{regularity_opt}.
\end{proof}
We adopt $r(\delta):=\sum_{j=1}^{n_X}1/\delta_j$, which satisfies the assumptions of Theorem~\ref{thm:risk_joint_solution} and is strictly convex.
Its monotone decrease encourages full use of the allowable risk budget, thereby avoiding unnecessarily conservative risk allocation, while its divergence at zero keeps individual allocations away from the boundary.
Strict convexity of the joint problem also supports stable numerical optimization by eliminating flat directions in risk allocation \cite{numopt}.

\section{Extension to learning-based EL models}\label{sec:exlin}
The linear-system formulation extends to learned EL models \cite{exlin,deep_sign_definite} through a linearizing change of variables.
Figure~\ref{fig:exlin} shows the model structure:
\begin{figure}[!t]
    \centering
    \includegraphics[width=\columnwidth]{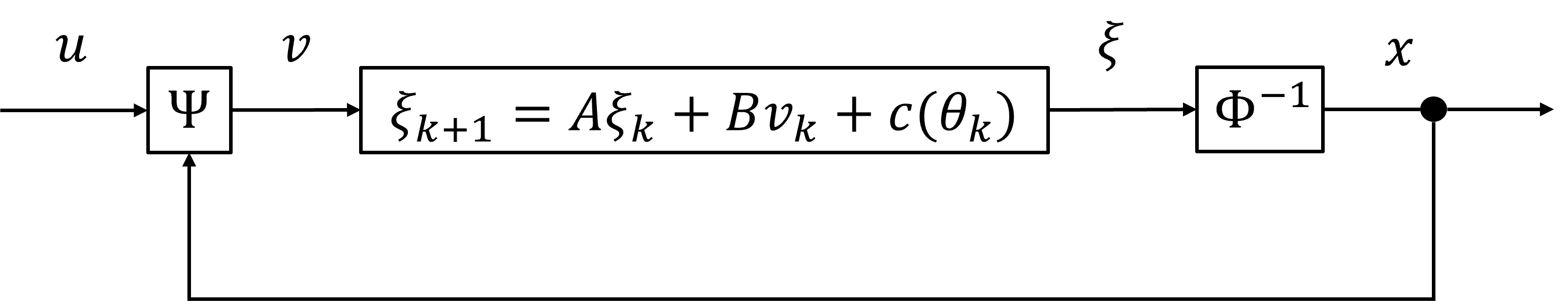}
    \caption{EL model}
    \label{fig:exlin}
\end{figure}
\begin{align}
    x_{k+1}=\Phi^{-1}(A\Phi(x_k)+B\Psi(u_k;x_k)+c(\theta_k)).
    \label{eq:state_eq_exlin}
\end{align}
Here $x_k\in\bbR^{n_x}$, $u_k\in\bbR^{n_u}$, and $\theta_k\in\bbR^{n_\theta}$.
The state and input transformations satisfy the following assumption, as in \cite{deep_sign_definite}.
\begin{ass}\label{ass:exlin_transformation}
The state transformation $\Phi:\bbR^{n_x}\to\bbR^{n_x}$ and the input transformation $\Psi:\bbR^{n_u}\times\bbR^{n_x}\to\bbR^{n_u}$ satisfy:
\begin{enumerate}
    \item $\Phi$ is elementwise and strictly increasing and satisfies $\Phi(0)=0$.
    \item For each fixed $x\in\bbR^{n_x}$, $\Psi(\cdot;x)$ is elementwise and strictly increasing and satisfies $\Psi(0;x)=0$.
\end{enumerate}
In particular, $\Phi$ and $\Psi(\cdot;x)$ are diffeomorphisms.
\end{ass}
The model is learned as described in \cite{exlin,deep_sign_definite}.
Following \cite{deep_sign_definite}, we formulate the tracking and effort costs in the coordinates $\xi:=\Phi(x)$ and $v:=\Psi(u;x)$:
\begin{align}
    &\min_v\ \sum_{k=0}^{N-1}\Bigl[
    w^\req\|\xi_{k+1}-\Phi(x^\req(\theta_{k+1}))\|^2+w^v\|v_k\|^2\nonumber\\
    &\hspace{4mm}+w^\soft\|\max(\xi_{k+1}-\Phi(\barx^\soft(\theta_{k+1})),0)\|^2\Bigr]
    \label{eq:opt_control_exlin_obj}\\
    &\st\xi_{k+1}=A\xi_k+Bv_k+c(\theta_k)\\
    &\Phi(\ubarx(\theta_{k+1}))\leq\xi_{k+1}\leq\Phi(\barx(\theta_{k+1}))\\
    &\Psi(\ubaru(\theta_k);\Phi^{-1}(\xi_k))\leq v_k\leq\Psi(\baru(\theta_k);\Phi^{-1}(\xi_k))
    \label{eq:opt_control_exlin_const}\\
    &k=0,\ldots,N-1.\nonumber
\end{align}
In MPC, the state arguments in the input bounds are fixed to the shifted state sequence predicted at the previous control step.
This approach is used in the numerical example.
With these arguments fixed, the transformed references and bounds depend only on the uncertain parameter and known data.
Introducing slack variables yields Problem~\eqref{eq:opt_control_lin_obj}--\eqref{eq:opt_control_lin_const}, in which the reference and state bounds are transformed by $\Phi$ and the input bounds by $\Psi$.
Thus, for stochastic $\theta$, Theorems~\ref{thm:risk_joint_convexity}--\ref{thm:risk_joint_solution} apply to the resulting transformed problem.

\section{Numerical example: Hybrid powertrain control}
\begin{figure*}[!t]
    \centering
    \includegraphics[width=\textwidth]{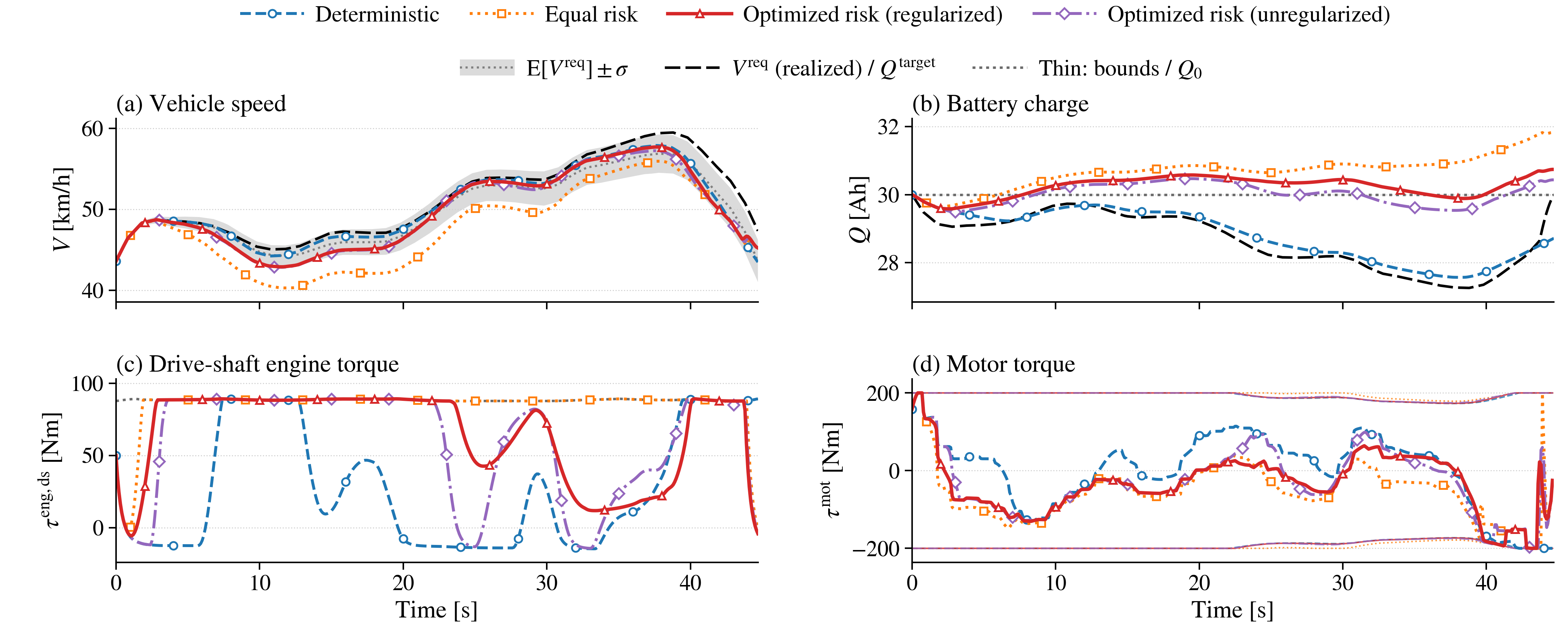}
    \caption{Closed-loop control responses.}
    \label{fig:control_responses}
\end{figure*}

\subsection{Controlled system and control specifications}
Following \cite{deep_sign_definite}, the state is defined as $x=[\tau^{\eng,\rmds},\ V,\ Q]$, representing the drive-shaft engine torque, vehicle speed, and battery charge, respectively.
The control input is defined as $u=[\tau^{\eng,\cmd},\ \tau^{\mot},\ \tau^{\brk}]$, corresponding to the commanded engine torque, motor torque, and brake torque, respectively.
The learned EL model is given by \eqref{eq:state_eq_exlin}, with a constant bias $c$.
A key engineering objective is charge sustaining, $Q_T\geq Q_0$, where $T$ is the terminal time of the full control interval.
This avoids obtaining a low short-term cost at the expense of the battery charge and preserves motor-use capability for the subsequent drive.

The physical control specifications are
\begin{align}
    &\min_{\tau^{\eng,\cmd},\,\tau^\mot,\,\tau^\brk}\ \sum_{k=0}^{N-1}
    \Bigl[w^\req(V_{k+1}-V_{k+1}^\req)^2\nonumber\\
    &\quad+w^\eng(\tau_k^{\eng,\cmd})^2+w^\mot(\tau_k^\mot)^2+w^\brk(\tau_k^\brk)^2\nonumber\\
    &\quad+w^\chg\bigl[\max(Q_{k+1}^\target(V_{k+1}^\req)-Q_{k+1},0)\bigr]^2\Bigr]
    \label{eq:powertrain_obj}\\
    &\st\text{\eqref{eq:state_eq_exlin} and}\nonumber\\
    &\tau_{k+1}^{\eng,\rmds}\leq\bartau^{\eng,\rmds}(V_{k+1}^\req)\\
    &\ubarQ\leq Q_{k+1}\leq\barQ\\
    &0\leq\tau_k^{\eng,\cmd}\\
    &\ubartau^\mot\leq\tau_k^\mot\leq\bartau^\mot,\quad
    \frac{\ubarP^\mot}{V_k^\req}\leq\tau_k^\mot\leq\frac{\barP^\mot}{V_k^\req}\\
    &0\leq\tau_k^\brk\leq\bartau^\brk\label{eq:powertrain_const}\\
    &k=0,\ldots,N-1.\nonumber
\end{align}
The objective represents speed tracking, torque effort, and the charge-target deficit.
The engine-torque limit is a requested-speed-dependent map derived from the NOx emission bound \cite{deep_sign_definite}.
The target-charge map $Q_k^\target(V_k^\req)$ is adapted from \cite{deep_sign_definite} to satisfy
\begin{align}
    Q_0^\target(V_0^\req)=Q_T^\target(V_T^\req)=Q_0.
    \label{eq:charge_target_endpoints}
\end{align}
These specifications are implemented in transformed coordinates as in Section~\ref{sec:exlin}, with $\theta:=V^\req$.

\subsection{Validation setup}
Table~\ref{tab:sim_param} lists the simulation settings.
The nominal requested speed is the WLTC Class 3b segment from $611.2$ to $655.9$~s, with the initial vehicle speed set to the segment's starting value.
This segment is selected to capture repeated acceleration and deceleration while maintaining equal initial and terminal nominal speeds and avoiding low speeds because the motor-power constraints contain inverse-speed terms.
The requested speed is obtained by adding the time integral of a rectangular acceleration perturbation to this nominal trajectory.
The full interval is divided into six nearly equal intervals, each assigned an independent zero-mean Gaussian perturbation with a standard deviation of $0.15$~km/h/s, and the initial speed perturbation is zero.
A single realized requested-speed trajectory, generated independently of the moment-estimation samples, is shared across all methods.
At each control step, the next requested speed is known, while subsequent speeds are characterized by their distribution.
Case (mv) of Assumption~\ref{ass:theta_distribution} is implemented using sample estimates of the required means and variances.

\begin{table}[!t]
    \caption{Simulation settings}\label{tab:sim_param}
    \centering
    \small
    \setlength{\tabcolsep}{3pt}
    \begin{tabular}{@{}>{\raggedright\arraybackslash}p{0.55\columnwidth}>{\raggedright\arraybackslash}p{0.40\columnwidth}@{}}
        \hline
        Item & Value\\
        \hline
        Sampling time & $0.1$ s\\
        Prediction horizon length $N$ & $10$\\
        Number of simulation steps & $447$ ($44.7$ s)\\
        Initial state $[\tau_0^{\eng,\rmds},V_0,Q_0]$ & $[50\ \mathrm{Nm},V_0^\req,30\ \mathrm{Ah}]$\\
        Speed-tracking weight $w^\req$ & $10$\\
        Engine-command weight $w^\eng$ & $1.0\times10^{-3}$\\
        Motor-torque weight $w^\mot$ & $5.0\times10^{-5}$\\
        Brake-torque weight $w^\brk$ & $1.0\times10^{-2}$\\
        Charge-deficit weight $w^\chg$ & $10$\\
        Risk-regularization weight $w^\delta$ & $1.0\times10^{-4}$\\
        NOx emission upper bound & $0.015$ mg/cycle\\
        Battery-charge bounds $\ubarQ,\barQ$ & $20,40$ Ah\\
        Motor-torque bounds $\ubartau^\mot,\bartau^\mot$ & $-200,200$ Nm\\
        Motor-power bounds $\ubarP^\mot,\barP^\mot$ & $-10000,10000$\\[-1pt]
        & $\mathrm{km/h\cdot Nm}$\\
        Brake-torque upper bound $\bartau^\brk$ & $50$ Nm\\
        Total allowable risk $\bardelta$ & $0.35$\\
        WLTC segment & Class 3b: $611.2$--$655.9$ s\\
        Acceleration-perturbation intervals & $6$\\
        Mean acceleration perturbation & $0$ km/h/s\\
        Acceleration-perturbation standard deviation & $0.15$ km/h/s\\
        Monte Carlo samples for moments & $1024$\\
        Independent samples for violation evaluation & $4096$\\
        \hline
    \end{tabular}
\end{table}

We compare four methods: The Deterministic method uses only the mean requested speed; Equal risk assigns $\bardelta/4$ to each of the four constraint types (engine upper, charge, motor lower, and motor upper) and then divides each share equally among its stochastic prediction times; Optimized risk (regularized) jointly optimizes the input and risk allocation with the weight in Table~\ref{tab:sim_param}; Optimized risk (unregularized) uses the same formulation with $w^\delta=0$.
All methods use the same physical constraints and remaining cost weights.
The corresponding optimization problems are solved using IPOPT \cite{ipopt}.

Solver time is the wall time recorded inside each solver call, summarized over the $447$ control steps by its mean, standard deviation, and maximum in milliseconds.
The closed-loop cost sums the implemented stage costs in transformed coordinates for the realized requested speed and applied input, excluding the risk-regularization term.
To estimate empirical violation probabilities, the optimal input sequence and charge-constraint slack are fixed at each control step.
For each constraint type, the fraction of $4096$ independent evaluation samples violating a constraint at least once within the prediction horizon is reported.
These samples are independent of those used for moment estimation.

\begin{figure*}[!t]
    \centering
    \includegraphics[width=\textwidth]{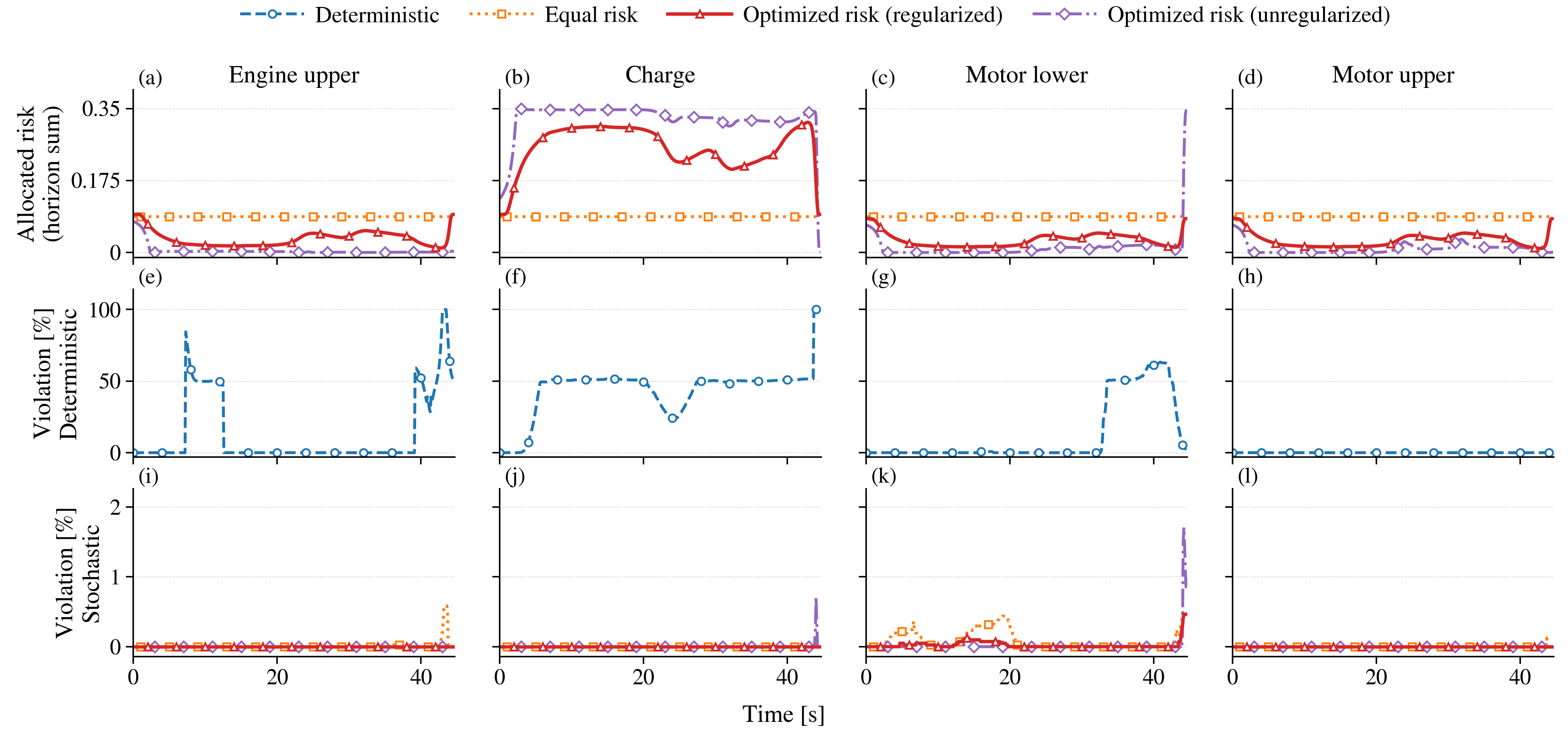}
    \caption{Horizon-summed risk allocations and empirical violation probabilities.}
    \label{fig:risk_and_violations}
\end{figure*}

\begin{table*}[!t]
    \caption{Solver time and closed-loop performance}\label{tab:performance}
    \centering
    \small
    \begin{tabular*}{\textwidth}{@{\extracolsep{\fill}}lcccc@{}}
        \hline
        Control method & \shortstack{Solver time [ms]\\Mean $\pm$ SD} & \shortstack{Maximum\\time [ms]} & \shortstack{Closed-loop\\cost (total)} & \shortstack{$Q_T-Q_0$\\{[Ah]}}\\
        \hline
        Deterministic & $15.407\pm0.829$ & $18.088$ & $28.6566$ & $-1.284$\\
        Equal risk & $16.062\pm1.095$ & $19.372$ & $122.4649$ & $+1.822$\\
        Optimized risk (regularized) & $21.999\pm2.945$ & $29.541$ & $31.6265$ & $+0.739$\\
        Optimized risk (unregularized) & $29.542\pm7.313$ & $109.605$ & $35.6778$ & $+0.432$\\
        \hline
    \end{tabular*}
\end{table*}

\subsection{Control responses}
Figure~\ref{fig:control_responses} compares the control responses.
When uncertainty in the requested speed is considered, the controller adjusts the engine–motor torque split to preserve battery charge for future operation.
The Equal risk method maintains a larger charge margin, but its conservative torque allocation degrades speed tracking.
Optimized risk uses the motor more flexibly while restoring the battery charge to at least its initial level at the end of the full interval.

\subsection{Risk allocation and performance evaluation}
Figure~\ref{fig:risk_and_violations} shows how the risk budget is allocated.
Optimized allocation assigns more risk to the charge constraint, reducing the excess charge margin associated with equal allocation.
With regularization, the allocations remain away from the boundary, while utilizing the total allowable risk, consistent with Theorem~\ref{thm:risk_joint_solution}.
Without regularization, some allocations approach zero and switch abruptly, particularly towards the end of the interval.
The independent evaluation of violation probabilities shows that accounting for uncertainty reduces violations relative to deterministic control; the low empirical rates also indicate conservatism of the sufficient conditions in this example.

Table~\ref{tab:performance} compares closed-loop cost and terminal charge.
Although deterministic MPC achieves a lower cost, it incurs a terminal charge deficit and therefore fails to satisfy the charge-sustaining objective.
All stochastic methods recover the initial charge.
Among the stochastic methods, optimized risk allocation achieves a lower closed-loop cost than equal risk allocation by avoiding an overly conservative charge margin.
Regularization also reduces both mean and maximum solver time compared with unregularized allocation.

\section{Conclusion}
This study formulates stochastic control under uncertain specifications as a strictly convex joint optimization of inputs and risk allocation.
The decreasing regularizer ensures the existence of an optimal solution and full use of the allowable risk, with uniqueness and continuity under the stated assumptions.
The extension of the proposed framework to learned EL models is validated through a hybrid powertrain control example comparing four control methods.
The results show that optimized risk allocation reduces excess charge margin and closed-loop cost relative to equal risk allocation, while regularization improves observed solver times.
Future work will investigate less conservative approximations compared with those based on Boole's and Cantelli's inequalities, evaluate broader WLTC segments with starts, stops, and gear changes, incorporate requested-speed uncertainty models learned from driving data, and perform hardware validation.

\appendix[Probability inequalities]
This section presents several probability inequalities used in this paper.
\subsection{Boole's inequality}
This subsection presents Boole's inequality \cite{probability_and_measure} and a related corollary.
\begin{lemma}[Boole's inequality]
For events $A_j\ (j\in\bbZ_m)$, the following holds:
\begin{align}
    \bbP\left[ \bigcup_{j=1}^m A_j \right] \leq \sum_{j=1}^m \bbP[A_j]
\end{align}
\end{lemma}
\begin{col}\label{cor:boole}
Let $X$ be an $n$-dimensional random vector, let $g_j:\bbR^n \to \bbR\ (j\in\bbZ_m)$ be functions, and let $\bardelta\in[0,1]$.
Assume that there exist $\delta_j\in[0,1]\ (j\in\bbZ_m)$ satisfying:
\begin{align}
    \sum_{j=1}^m \delta_j \leq \bardelta,\ \bbP[g_j(X)\leq 0] \geq 1-\delta_j\ (j\in\bbZ_m).
\end{align}
Then:
\begin{align}
    \bbP\left[ g_j(X) \leq 0\ (j\in\bbZ_m) \right] \geq 1 - \bardelta.
\end{align}
\end{col}

\subsection{Cantelli's inequality}
This subsection presents Cantelli's inequality \cite{concentration_inequalities} and a related corollary.
\begin{lemma}[Cantelli's inequality]
For a one-dimensional random variable $X$ and a real number $t> 0$, the following holds:
\begin{align}
    \bbP[ X-\bbE[X] \geq t ] \leq \dfrac{\bbV[X]}{\bbV[X] + t^2}.
\end{align}
\end{lemma}
\begin{col}\label{cor:cantelli}
Let $X$ be a one-dimensional random variable, let $y\ (\neq \bbE[X])$ be a real number, and let $\delta\in(0,1]$.
Assume that:
\begin{align}
    \bbE[X] + \sqrt{ \dfrac{1-\delta}{\delta} \bbV[X] } \leq y \label{eq:cantelli_condition}
\end{align}
Then $\bbP[X \leq y] \geq 1-\delta$ holds.
\end{col}

\subsection{Hoeffding's inequality}
This subsection presents Hoeffding's inequality \cite{concentration_inequalities} (one-dimensional version) and a corollary.
\begin{lemma}[Hoeffding's inequality]
For a one-dimensional random variable $X$ that satisfies $X\in[L^X, U^X]$ almost surely and for a real number $t>0$, the following holds:
\begin{align}
    \bbP[ X-\bbE[X] \geq t ] \leq \exp\left( -\dfrac{2t^2}{(U^X - L^X)^2} \right).
\end{align}
\end{lemma}
\begin{col}\label{cor:hoeffding}
Let $X$ be a one-dimensional random variable that satisfies $X\in[L^X, U^X]$ almost surely, and let $y\ (\neq \bbE[X])$ be a real number and $\delta\in(0,1]$.
Assume that:
\begin{align}
    \bbE[X] + \dfrac{U^X - L^X}{\sqrt{2}} \sqrt{ -\log(\delta) } \leq y \label{eq:hoeffding_condition}
\end{align}
Then $\bbP[X \leq y] \geq 1-\delta$ holds.
\end{col}

\bibliographystyle{IEEEtran}
\bibliography{references}

\end{document}